\documentclass{article}

\usepackage{PRIMEarxiv}

\usepackage[utf8]{inputenc} % allow utf-8 input
\usepackage[T1]{fontenc}    % use 8-bit T1 fonts
\usepackage{hyperref}       % hyperlinks
\usepackage{url}            % simple URL typesetting
\usepackage{booktabs}       % professional-quality tables
\usepackage{amsfonts}       % blackboard math symbols
\usepackage{nicefrac}       % compact symbols for 1/2, etc.
\usepackage{microtype}      % microtypography
\usepackage{lipsum}
\usepackage{fancyhdr}       % header
\usepackage{graphicx}       % graphics
\graphicspath{{media/}}     % organize your images and other figures under media/ folder

\title{Poleval: A Python package for HAXPES analysis}
\author{
  Robin Yoël Engel\\
  Wallenberg Initiative Materials Science for Sustainability,\\
  Chemical Physics, Department of Physics,  \\
  Stockholm University, Roslagstullsbacken 21, \\
  114 21 Stockholm\\
  \texttt{robin.engel@fysik.su.se} \\
   \And
  Patrick Lömker \\
  Chemical Physics, Department of Physics,  \\
  Stockholm University, Roslagstullsbacken 21, \\
  114 21 Stockholm\\
  \texttt{robin.engel@fysik.su.se} \\
}

\begin{document}
\maketitle

\begin{abstract}
POLEVAL provides a software toolbox for collaborative, persistent and reproducible analysis of XPS experiments. It allows to treat, analyse and visualise the results of an extended experimental campaign in a single python notebook in a consistent manner. Managing experimental data in adequate objects enables experimentalists to process and analyse measurements in very few lines of code, so as to provide decision aids through online data analysis during e.g. beamtime experiments. The persistent and self-documentary style of the notebook-based analysis allows for easy communication of intermediate results and enables progressive refinements into publishable figures or exporting the results to other programs.

The toolbox facilitates various routines for data treatment (normalization, cropping, etc.) and aggregation of spectra into groups to analyse trends. It also enables quantitative analysis with three major functions:

First, normalization to the photoionization cross-section and probability of emission into the analyser cone allows for quantitative comparisons between intensities from different core levels. The integrated \texttt{haxquantpy} package allows easy retrieval of literature values \cite{trzhaskovskayaDiracFockPhotoionization2018} for this purpose. 

Second, an extensive fitting functionality is implemented to treat groups of spectra together, rather than spectrum-by-spectrum. This grouping allows reinforcing the fit algorithm with prior knowledge, such as the equivalence of peak widths or positions between spectra, which enables for more consistent, and importantly, more confident fit results for sets of potentially noisy spectra.

Third, a simple formalism to estimate the thickness of adsorbate layers based on the ratio between the substrate's and adsorbate's XPS signal is implemented.\end{abstract}

% keywords can be removed

\section{Statement of Need}
XPS spectra are often analysed using home developed tools to perform the analysis. Here, results are reported by specifying spectral shapes and background treatment, but often without a public discussion or publication of the actual procedures, or with the use of commercial software (Casa XPS, UniFit, ...).  

We find that common open-source source algorithms, including their publication, discussion and common development can be extremely beneficial for both the efficiency and scientific rigour of experimental work.

While some open-source software for XPS is available (i.e. \cite{nakajimaLG4X2024} or \cite{stansburyPyARPESAnalysisFramework2020}), published software applicable to the HAXPES regime and a with a focus on code efficiency is so far lacking.

Our work is inspired by the special environment of synchrotron-based research, which necessitates a frequent information hand-over and near-online data analysis. This motivates the choice of a Jupyter notebook where a documented form of the proceedings (markdown descriptions, headers together with spectra and analysis results) of the last shifts is inherently generated and can be worked on continually. Complex datasets often probe a certain dimension and vary one parameter over sets of spectra. Consequently POLEVAL enables quick plotting with many functions being able to get a single line of code to combine a complex data set, while also enabling a smooth transition into non-standard in-depth analysis within the same framework.

The next step after loading and manipulating data is analysing the spectral contributions. The ability to enforce consistent (or even equal) peak shape parameters over entire measurement series is essential, especially in chemically sensitive applications \cite{majorPracticalGuideCurve2020}, yet no published software so far provides this functionality, to the best of the authors knowledge. 

The big strength of XPS is that it is quantitative, a spectrum can be related to the absolute abundance of chemical species in the sample surface-near region.  Experiments in the HAXPES regime using grazing incidence X-rays necessitate a modelling of the X-ray penetration length in combination with the electron emission pathway. This feature is implemented in the tool such that, given the incidence angle and an average adsorbate composition, sample surface coverages can be quantified in terms of mono-layers.

\section{Implementation}
\begin{figure}
    \centering
    \includegraphics[width=1\linewidth]{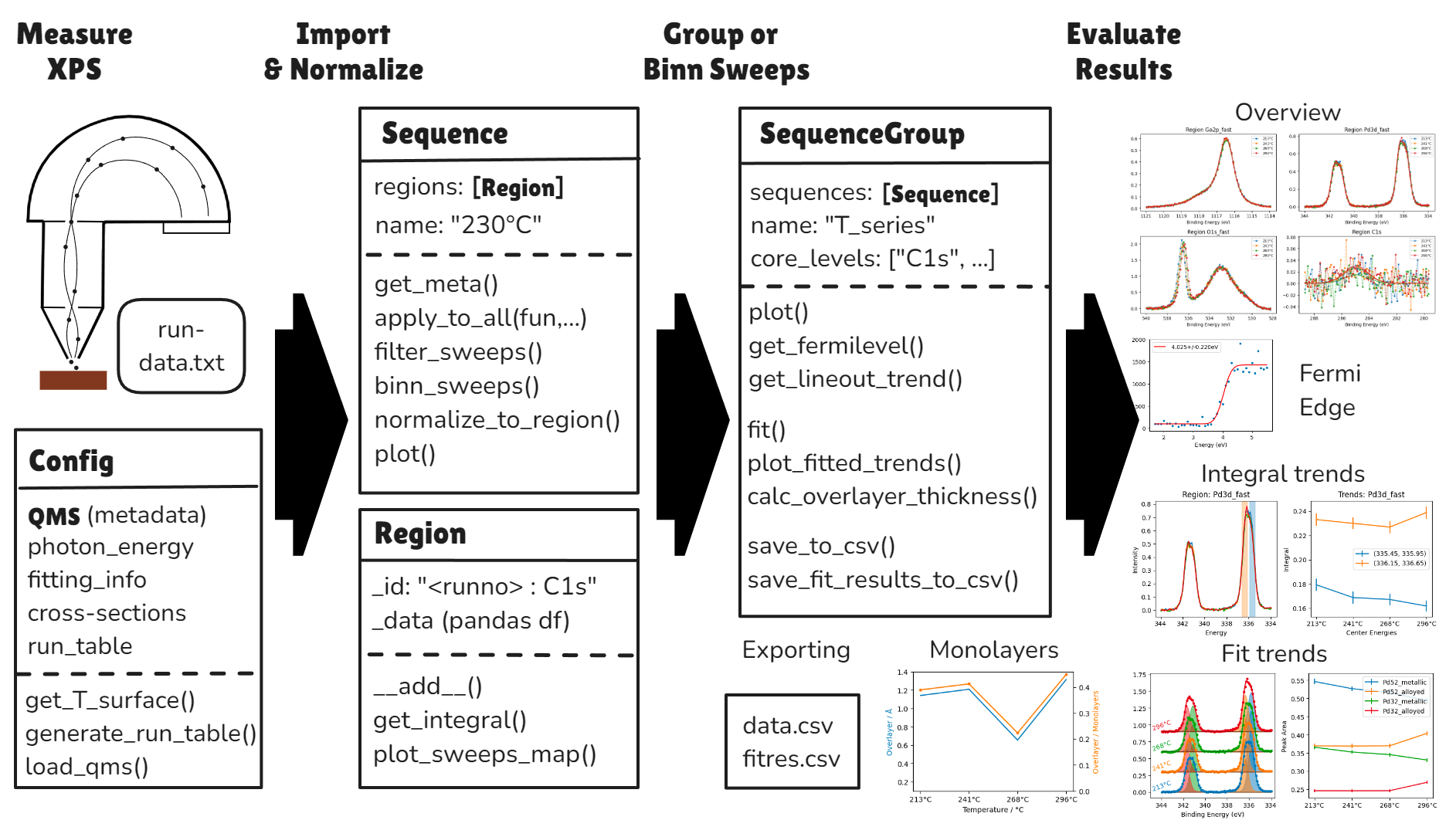}
    \caption{Data pipeline using the poleval package, showing the key classes and and their most important properties and methods as well as typical result plots.}
    \label{fig: Implementation Figure}
\end{figure}
The principal functions of POLEVAL are illustrated in \autoref{fig: Implementation Figure}. Spectra and acquisition settings are read from text files generated by the SES software which controls the Scienta XPS analyser. While XPS data from other sources (e.g. SPECS) can be read, the SES text files are tested most.

Spectra are organised in \texttt{Region} objects that represent electronic core levels. Each \texttt{Region} object represents a single core level spectrum and retains information of the individual sweeps as well as the normalizations applied in the import step.

An experiment typically compares sets of regions, which are recorded together in an interleaved manner, \texttt{Region} objects are thus organised into \texttt{Sequence} objects, which correspond to a loaded file.  Key scientific insights often emerge from analysing trends - be it dependent on time, temperature, gas pressure or other experimental variables. Such comparisons are enabled by the \texttt{SequenceGroup} class.

A \texttt{SequenceGroup} can be generated either from a list of separately imported and treated \texttt{Sequence} objects, or if applicable, by breaking up and binning the sweeps contained in the \texttt{Regions} of a single \texttt{Sequence}. Each of these classes contains rich functionality to facilitate the manipulation and intermediate visualization of data, while also maintaining reproducibility of each data treatment step taken. A \texttt{Config} object is used to store global settings, such as the photon energy, but also fit models for each core level and metadata in form of a \texttt{QMS} object, which can be used for binning or generating a run table as an experiment overview. 

The peak fitting algorithm is based on the \texttt{lmfit} package, but operates on all \texttt{Region}s of the same core level within one \texttt{SequenceGroup}, and allows for dependency constraints between fit parameters. Common applications are to enforce a common (yet variable) peak width across multiple peaks and spectra, or defining a fixed ratio or distance between peak parameters in one spectrum. Defining such constraints by imposing prior knowledge helps to restrict the information content that must be extracted from potentially noisy data, and thus reduces the uncertainty of the fit results.

Similarly, the estimation of the adsorbate overlayer thickness requires a set of parameters, such as the lattice and atomic constants of each material involved, to be defined explicitly before an estimate can be computed. The mathematic model is laid out below.

\section{Overlayer Thickness Model}

\begin{figure}[h]
    \centering
    \includegraphics[width=0.5\linewidth]{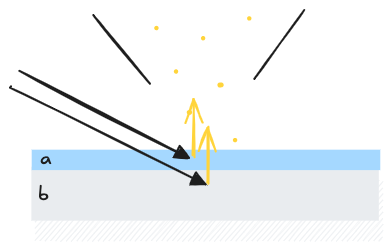}
    \caption{Schematic of the modeled geometry: From the left, parallel X-rays impinge at a small incidence angle onto the sample, which consists of a bulk material b (possibly alloyed with a second element c), but has a thin adsorbate layer a on top. Both emit photo-electrons into the XPS analyser cone. The signal from the bulk is attenuated more due to the absorption in the adsorbates, enabling a quantification by comparing the signal levels.}
    \label{fig: geometry}
\end{figure}

We use a model for estimating the adsorbate coverage which assumes a single homogeneous flat adsorbate (indexed \textit{a}) on top of an equally flat and homogeneous semi-infinite bulk substrate (potentially of two constituents, "double substrate", indexed \textit{b} and \textit{c}). This geometry is sketched in Figure \ref{fig: geometry}. The model is meant to give estimates for thicknesses on the Angstrom scale, and does not account for surface roughness, as well as reflections or standing-wave effects. The rationale for this is that the differences in X-ray intensity, which adsorbates should experience, are negligible because they are on 10\,nm scales and adsorbates on the 0.2\,nm scale.

The variables are listed in \autoref{tab:variables}, where the subscripts $a$ and $b$ are used for the absorbate and bulk, respectively, and $A,B$ for the respective core levels. The effective probing depth is:

\begin{equation}
    \frac{1}{\lambda'} = \frac{1}{\lambda_p} + \frac{1}{\lambda_e}
\end{equation}

For a single-layered system, we can describe the detectable signal strength as:
\begin{equation}
S=I_0 \rho \sigma \lambda'
\end{equation}
Considering a bilayer system, the integration has to be performed explicitly. 
The signal from the adsorbate layer becomes:
\begin{equation}
S_a = I_0 \rho_a \sigma_a \int_0^{d}e^{-\frac{z}{\lambda_{a,A}'}} dz = I_0 \rho_a \sigma_a \lambda_{a,A}' \left( 1-e^{-\frac{d}{\lambda_{a,A}'}} \right)
\end{equation}
While the signal of the bulk is attenuated by two additional exponential terms representing the attenuation of the incident X-rays and the scattering of the emitted electrons in the adsorbate layer, respectively.
\begin{equation}
S_b = I_0 e^{-\frac{d}{\lambda_{p,a}}} \rho_b \sigma_b \lambda_{b,B}' e^{-\frac{d}{\lambda_{e,a,B}}}
\end{equation}

Since the precise sensitivity of the experimental setup is challenging to characterise, the more reliable experimental quantity is the ratio of adsorbate and bulk signal. This becomes:
\begin{equation}
\frac{S_a}{S_b} = \frac{I_0 \rho_a \sigma_a \lambda_{a,A}' \left( 1-e^{-\frac{d}{\lambda_{a,A}'}} \right)}{I_0 \rho_b \sigma_b \lambda_{b,B}' e^{-\frac{d}{\lambda_{a,B}'}}}
\end{equation}
We cannot solve this equation analytically, but instead employ a root finding algorithm (scipy.optimize.fsolve) to solve for \textit{d}.
This is implemented in the \texttt{poleval.overlayer\_thickness.find\_thickness(...)} function.

\subsection{Simplified form}

Under the approximation that the effective probe depth in the adsorbate are equal for both core levels $\lambda_{a,A}' \simeq \lambda_{a,B}'$, we can write out an analytically closed form for $d$:
\begin{equation}
d = \lambda _a' log \left( \frac{S_a \lambda _b' \rho _b \sigma _b}{S_b \lambda _a' \rho _a \sigma _a} + 1 \right) 
\end{equation}
This is implemented in \texttt{poleval.overlayer\_thickness.simplified\_adsorbate\_thickness(...)} and is also used as an initial guess for the root finding algorithm.

\subsection{Composite substrate}
If two substrate core levels have been measured, e.g. because the substrate is an alloy, one may use both of those signals for a more accurate thickness estimate. This is described by this formula:
\begin{equation}
\frac{S_a}{S_b + S_c} = \frac{\rho_a \sigma_a \lambda_{a,A}' (1-e^{-d/\lambda_{a,A}'})}{\rho_b \sigma_b \lambda_{b,B}' e^{-d/\lambda_{a,B}'}+\rho_c \sigma_c \lambda_{c,C}' e^{-d/\lambda_{a,C}'}},
\end{equation}
where the index $c$ describes the second substrate element (and core level $C$), and implemented in the \texttt{poleval.overlayer\_thickness.find\_thickness\_double\_substrate(...)} function.

\begin{table}[h]
\centering
\begin{tabular}{ll}
\textbf{Symbol} & \textbf{Description} \\
\hline
$\lambda_p$     & X-ray penetration length (at the given incidence angle) [m] \\
$\lambda_e$     & Electron escape depth [m] \\
$\lambda'$      & Effective probing depth [m] \\
$I_0$           & Incident intensity \\
$\rho$          & Density [atoms m$^{-3}$] \\
$\sigma$        & Photo-ionization cross-section [m$^{-2}$] \\
$d$             & Adsorbate layer thickness [m] \\
$z$             & Sample depth \\
$S$             & Signal intensity from photoelectrons leaving the sample \\
                & (not accounting for instrumental functions) \\
\hline
\end{tabular}
\caption{List of variables.}
\label{tab:variables}
\end{table}

\section{Availability}
The software is freely available at \url{https://gitlab.fysik.su.se/operando-catalysis-spectroscopy/polariseval/-/tree/joss_paper} under an MIT licence.

\section*{Acknowledgments}
This work was partially supported by the Wallenberg Initiative Materials Science for Sustainability (WISE) funded by the Knut and Alice Wallenberg Foundation.
We acknowledge contributions from Mikhail Shipilin to specqp \cite{shipilinSPECQPStandsSPECtroscopy2025} that we used as the starting point to this project.

%Bibliography
\bibliographystyle{IEEEtran}  
\bibliography{references}

% Generated by IEEEtran.bst, version: 1.14 (2015/08/26)
\begin{thebibliography}{1}
\providecommand{\url}[1]{#1}
\csname url@samestyle\endcsname
\providecommand{\newblock}{\relax}
\providecommand{\bibinfo}[2]{#2}
\providecommand{\BIBentrySTDinterwordspacing}{\spaceskip=0pt\relax}
\providecommand{\BIBentryALTinterwordstretchfactor}{4}
\providecommand{\BIBentryALTinterwordspacing}{\spaceskip=\fontdimen2\font plus
\BIBentryALTinterwordstretchfactor\fontdimen3\font minus \fontdimen4\font\relax}
\providecommand{\BIBforeignlanguage}[2]{{%
\expandafter\ifx\csname l@#1\endcsname\relax
\typeout{** WARNING: IEEEtran.bst: No hyphenation pattern has been}%
\typeout{** loaded for the language `#1'. Using the pattern for}%
\typeout{** the default language instead.}%
\else
\language=\csname l@#1\endcsname
\fi
#2}}
\providecommand{\BIBdecl}{\relax}
\BIBdecl

\bibitem{trzhaskovskayaDiracFockPhotoionization2018}
\BIBentryALTinterwordspacing
M.~Trzhaskovskaya and V.~Yarzhemsky, ``Dirac–{{Fock}} photoionization parameters for {{HAXPES}} applications,'' \emph{Atomic Data and Nuclear Data Tables}, vol. 119, pp. 99--174, 01 2018. [Online]. Available: \url{https://linkinghub.elsevier.com/retrieve/pii/S0092640X16300596}
\BIBentrySTDinterwordspacing

\bibitem{nakajimaLG4X2024}
\BIBentryALTinterwordspacing
H.~Nakajima. {{LG4X}}. Zenodo. [Online]. Available: \url{https://zenodo.org/doi/10.5281/zenodo.10477914}
\BIBentrySTDinterwordspacing

\bibitem{stansburyPyARPESAnalysisFramework2020}
\BIBentryALTinterwordspacing
C.~Stansbury and A.~Lanzara, ``{{PyARPES}}: {{An}} analysis framework for multimodal angle-resolved photoemission spectroscopies,'' \emph{SoftwareX}, vol.~11, p. 100472, 1 2020. [Online]. Available: \url{https://www.sciencedirect.com/science/article/pii/S2352711019301633}
\BIBentrySTDinterwordspacing

\bibitem{majorPracticalGuideCurve2020}
\BIBentryALTinterwordspacing
G.~H. Major, N.~Fairley, P.~M.~A. Sherwood, M.~R. Linford, J.~Terry, V.~Fernandez, and K.~Artyushkova, ``Practical guide for curve fitting in x-ray photoelectron spectroscopy,'' \emph{Journal of Vacuum Science \& Technology A: Vacuum, Surfaces, and Films}, vol.~38, no.~6, p. 061203, 12 2020. [Online]. Available: \url{https://pubs.aip.org/jva/article/38/6/061203/1023652/Practical-guide-for-curve-fitting-in-x-ray}
\BIBentrySTDinterwordspacing

\bibitem{shipilinSPECQPStandsSPECtroscopy2025}
\BIBentryALTinterwordspacing
M.~Shipilin. {{SPECQP}} stands for {{SPECtroscopy Quick Peak}}. [Online]. Available: \url{https://github.com/Shipilin/specqp}
\BIBentrySTDinterwordspacing

\end{thebibliography}

\end{document}